Yan Shvartzshnaider*, Zvonimir Pavlinovic, Thomas Wies, Lakshminarayanan Subramanian, Prateek Mittal , and Helen Nissenbaum

# The VACCINE Framework for Building DLP Systems

**Abstract:** Conventional Data Leakage Prevention (DLP) systems suffer from the following major drawback: Privacy policies that define what constitutes data leakage cannot be seamlessly defined and enforced across heterogeneous forms of communication. Administrators have the dual burden of: (1) manually self-interpreting policies from handbooks to specify rules (which is error-prone); (2) extracting relevant information flows from heterogeneous communication protocols and enforcing policies to determine which flows should be admissible. To address these issues, we present the Verifiable and ACtionable Contextual Integrity Norms Engine (VACCINE), a framework for building adaptable and modular DLP systems. VACCINE relies on (1) the theory of contextual integrity to provide an abstraction layer suitable for specifying reusable protocol-agnostic leakage prevention rules and (2) programming language techniques to check these rules against correctness properties and to enforce them faithfully within a DLP system implementation. We applied VACCINE to the Family Educational Rights and Privacy Act and Enron Corporation privacy regulations. We show that by using contextual integrity in conjunction with verification techniques, we can effectively create reusable privacy rules with specific correctness guarantees, and check the integrity of information flows against these rules. Our experiments in emulated enterprise settings indicate that VACCINE improves over current DLP system design approaches and can be deployed in enterprises involving tens of thousands of actors.

**\*Corresponding Author: Yan Shvartzshnaider:** NYU/Princeton, ys63@nyu.edu
**Zvonimir Pavlinovic:** NYU, zvonimir@cs.nyu.edu
**Thomas Wies:** NYU, wies@cs.nyu.edu
**Lakshminarayanan Subramanian:** NYU, lakshmi@cs.nyu.edu
**Prateek Mittal :** Princeton, pmittal@princeton.edu
**Helen Nissenbaum:** NYU, hfn1@nyu.edu

## 1 Introduction

Enterprises rely on Data Leakage Prevention (DLP) systems to prevent accidental or unintentional flow of sensitive information such as credit card and other business data to unauthorized entities [1]. Most deployed DLP products, such as [2–6], include a centralized administrative system that takes domain-specific rules written by administrators as input to enforce the specific DLP policy of the enterprise [1, 5, 6]. We argue that the design of this central component of state-of-the-art DLP systems has several shortcomings and propose a new framework to address these.

The first shortcoming concerns usability. On the one hand, the process of applying policies "*[does not] fit the low technical expertise of many policy authors*" [7]. On the other hand, administrators of rule-based DLP systems have to manually implement the rules using privacy handbooks written by lawyers.

Existing DLP systems do not allow the rules to be expressed at the right level of abstraction. In most systems, policies are reduced to a ruleset of regular expressions, templates, keywords, or patterns [8], which conflates two aspects of DLP that should be addressed separately: 1) how to extract information flows from communication traffic, and 2) how to enforce the leakage prevention rules that determine which extracted flows are admissible. For example, policy authors who specify DLP rules should not have to think about how to extract a credit card number or other piece of sensitive information from a particular source of communication, as this might require additional technical expertise. Instead, they should be enabled to specify rules on a more abstract level that only concerns the information flows implicit in these communications. Moreover, existing DLP systems do not provide support tools that enable administrators to check the correctness or completeness of the specified rules. It is therefore not surprising that the process of translating policy handbooks to DLP rules is often error-prone [9] and relying on incomplete default templates. Companies often rely on third-party providers such as Nucleuz "to develop a cus-tom DLP policy to detect and protect [the] company's specific data and assets" [10].

Second, there is a need for "*developing flexible policy frameworks that can deploy the same logical policy to heterogenous devices or systems with incompatible abstractions*" [7]. However, conflating flow extraction and leakage prevention is in direct conflict with this principle and unduly limits the effectiveness and applicability of existing DLP systems. Rules defined by, say, regular expressions target unstructured data and are therefore less suited for structured forms of communication such as web forms and directory service maps [11]. The leakage prevention rules should abstract from the specific type of communication from which information flows are extracted and apply to both structured and unstructured data sources alike.

In response, this paper presents Verifiable and ACtionable Contextual Integrity Norms Engine (VACCINE), a new framework for building powerful and modular DLP systems. VACCINE casts data leakage prevention as a privacy problem expressed in the theory of Contextual Integrity (CI) [12]. More precisely, VACCINE uses CI to model the information flows in an enterprise and the notion of data leakage, and to express privacy policies as actionable (mechanized) rules for preventing these leakages using standard programming language techniques. DLP systems that follow the VACCINE framework do not have to assume any particular class of application domains or communication exchange formats. The specification of DLP rules and checking information flows for data leakages is done on the abstract level defined in terms of a few simple concepts of CI theory.

Our evaluation of the proposed framework examines the question of how to effectively design an enterprise DLP system using VACCINE given a handbook that outlines the privacy and confidentiality policies. As a case study, we evaluate VACCINE on the Family Educational Rights and Privacy Act and Enron Corporation privacy regulations. We show that we can create privacy rules with certain correctness guarantees and effectively check for the integrity of information flows, i.e., data leakages, against these regulations. Our experiments indicate that VACCINE can effectively and efficiently process privacy-sensitive information exchanges between tens of thousands of actors in emulated enterprise settings.

## 2 Design

In this section we start by stating the threat model assumptions and elaborate on the design goals behind VACCINE. We then provide a brief overview of contextual integrity and discuss how it is implemented in VACCINE framework.

### 2.1 Threat Model Assumptions

In our design of the VACCINE framework we make the following main threat model assumptions faced by the existing DLP systems:

1. Modern DLP solutions are maintained by system admins or other IT personal who are in charge of manually mapping the relevant governing policies and regulation into a set of DLP rules.
2. DLP systems handle information exchanges across heterogeneous systems with varied application logics and therefore are subject to logical misinterpretations due an oversight by the admin personal.
3. The enforced DLP policies can significantly deviate from privacy expectations of the system end-user, which can lead to unintentional breach.
4. DLP system is designed to prevent accidental information leakage due to an unintentional breach of a regulation or policy.
5. A typical end user is not engaged in a malicious behavior to circumvent the existing DLP measures.

We now describe the key building blocks behind the VACCINE framework.

### 2.2 Design goals

VACCINE treats information leakage prevention not as a security measure (blocking access to sensitive content) but rather as a privacy logic verification problem (ensuring that information flows are consistent with the prescribed policies). By using a well-known privacy framework as a specification of *allowed* information flows which are then expressed in a separate form of declarative logic, VACCINE decouples the processes of defining and enforcing the policy rules. More specifically, VACCINE aims to achieve a) *usability* by allowing non-technical personnel to devise privacy policies that can be directly enforced by system administrators, b) *adaptability* by defining DLP rules in a way that abstracts from flow extraction and allowing one rule en-forcement engine to be combined with flow extractors for different types of communications, and c) *modularity* by separating the processes of capturing information flows and enforcing DLP rules and, consequently, enabling engineering and research efforts to focus on enhancing each aspect in isolation. We now explain in more details the manner in which VACCINE achieves its design goals.

**Usability** VACCINE uses an intermediate abstraction in the form of the CI privacy framework, explained in Section 2.3, between the enforced DLP rules and the logic prescribed by the privacy handbooks. Table 1 shows the mapping the original text of the policy into a CI representation and eventually into VACCINE's enforceable and verifiable rules, columns 1-3 respectively.

This removes the need for the system administrators to self-interpret the original text of privacy policies (column 1) into rules enforced by the modern DLP systems (column 4). Such a burdensome process can lead to misinterpretations and inconsistent assumptions, as we discovered in our own evaluation (Section 4.2).

With VACCINE the legal and Chief Privacy Officer (CPO) departments (we refer to them as a DLP team throughout the paper) are able to check policies before deploying them. The rules are expressed in terms of the theory of contextual integrity which in turn has known formulations in formal logics [13, 14] and as a consequence, the devised rules can be checked for various correctness properties using formal methods. This way, many bugs in the implementation of a specific DLP policy can be caught before deployment. This allows the administrator to map the devised policies into VACCINE rules by just focusing on system and enterprise information essential.

**Adaptability** The VACCINE rules abstract away from low-level system details, they can be enforced in heterogenous environments. Hence, the same rules can be integrated into existing applications running on different devices and relying on different communication exchanges. For example, the policy in Table 1: *Faculty members (FM) do not have access to student academic records (AR) unless they have a "legitimate educational interest,"* which in current DLP system requires separate policies and different enforcement mechanisms for different platforms, such as for email communication and for cloud access (e.g., MS Exchange [6] and MS Sharepoint/One Drive [15], respectively). The VACCINE policy can be used across other communication platforms. The administrators just need to specify one set of rules that govern leakage prevention for all types of enterprise communication (emails, tweets, web forms, etc.)

**Modularity** The functionality for preventing data leakages works on the abstract level defined in terms of concepts of contextual integrity. We emphasize the modular nature of this design. As shown in Figure 1, in VACCINE, the CI flow checker, which is responsible for ensuring the consistency of flows, is decoupled from the rest of the application that provides conversion from concrete communication exchanges to sequences of CI flows. The checker, is parameterized by privacy rules and is hosted on the server side, unaware of the application layer innerworkings and communication exchange formats.

Not only does that this allows for widespread integration into existing systems, it also provides the application with additional informational obfuscation layer. Using policies in Table 1 as an example; with VACCINE, there is no need to reveal the actual personal observations or academic records, as required by traditional DLP solutions (see the Condition clause in column 4), in order to check if the information flow adheres to FERPA policies. Sensitive information is obfuscated by the application through the using CI language, which, as we discuss in Section 2.3, only reveals the type of information being conveyed not its actual value. The responsibility of such applications is to implement functionality that extracts information flows from communication exchanges and passes the flows to the DLP component enforcing the rules.

## 2.3 Contextual Integrity Overview

In our work, we adopt the theory of contextual integrity (CI) as our underlying conception of privacy. The power of CI comes from its simple, yet not trivial way of capturing and subsequently reasoning about the governing privacy norms in a given context. More specifically, unlike other privacy frameworks, CI postulates that informational privacy is not all about secrecy (blocking information) [17] or control [18] but about the appropriateness of information flow within a particular context. Appropriateness of flow means flow that is compliant with contextual norms governing informational flows. We now describe CI in more detail.

**CI model.** The building blocks of CI are *actors*, *attributes* and *transmission principles*. Actors are concrete participants involved in an information exchange. An actor can be a particular person or institution. As a participant, actors can possess roles that describe the capacity in which they function in a given context such as doctors or teachers. An attribute is a piece of information being exchanged, such as a grade, a medical record, etc. Finally, a *transmission principle* reflects the constraints associated with a given information exchange (e.g., under strict confidentiality).

| Policy | Contextual Integrity | VACCINE | DLP solution |
|---|---|---|---|
| Personal observations (PO) made by the faculty member (FM) about a student may not be disclosed without the student's consent. | ($FM_{sndr}$, $Anyone_{rcp}$, $Student_{subj}$, $PO_{attr}$, with permission from the student$_{TP}$) | allowed($FERPA_{Ctx}$, $FM_{sndr}$, $Anyone_{rcp}$, $Student_{subj}$, $PO_{attr}$) explicit_permission($Student_{subj}$, $FM_{sndr}$, $Anyone_{rcp}$, $PO_{attr}$) | **Conditions:** a) PO in the content b) sender is a Faculty Member **Action:** a) stop the email b) display warning c) prompt permission to forward it to Student for approval |
| Faculty members (FM) do not have access to student academic records (AR) unless they have a "legitimate educational interest" (lei). | (System, $FM_{rcp}$, $Student_{subj}$, $ERec_{attr}$, with legitimate educational interest$_{TP}$) | allowed($FERPA_{Ctx}$, $System_{sndr}$, $FM_{rcp}$, $Student_{subj}$, $AR_{attr}$) lei($FM_{rcp}$, $Student_{subj}$, $AR_{attr}$) | **Conditions:** a) content includes academic records **Exception:** Sender is marked as Faculty Member and has a role which grants him lei access, e.g., as a advisor of the student **Action:** Stop content from being sent |

**Table 1.** Examples on how policies are captured using CI and implemented using VACCINE in comparison to a generic rule-based DLP solution. In the DLP solution column the rules comprise conditions, exceptions, actions, properties: [16]. **Conditions:** filter messages that require attention. Conditions comprise predicated on senders, recipients, subject, body, headers and attachments. **Exceptions** filter messages that are exempt from any conditional constrains. **Actions** specify what to do with the messages that matched by conditions and are not exempt by exceptions, e.g., block the message, redirecting, editing original content and recipients set. **Properties** specify parameters under which the rule is operational. e.g., on what date the rule is activated, mode of operation (test or live), and so forth. Note, although we used the terminology from Microsoft Exchange in column 4, all larger DLP software provider offer similar semantics to express DLP rules.

**Information flows.** One appealing aspect of CI is that it provides a simple and clear definition of *information flows*. An information flow is a tuple

`<sender, recipient, subject, attribute>`

Sender, recipient, and subject are actors. A flow represents an atomic unit of an information exchange where `sender` is sending `attribute` that refers to `subject`, to `recipient`. From the CI point of view, communication exchanges are simply sequences of flows defined in this manner.

*Contextual Informational Norms* specify what flows are allowed in a given privacy context constrained by *transmission principles*. An informational norm can also be described as a tuple:

`<Sender, Recipient, Subject, Attribute, transmission principle>`

Here `Sender`, `Recipient`, and `Subject` refer to roles and `Attribute` to a particular type of attributes. For instance, in the educational setting, where information exchange between actors is regulated by the FERPA policies, *students* (the role of the sender and subject) would be allowed to tell their *parents* (the role of the recipient) about their *grades* (the type of the attribute). The corresponding informational norm could specify the transmission principle of *reciprocity*, as something that is expected to happen. However, when a parent asks for the school to provide the student's record, i.e., originate an information flow where the school, acts as a sender, the parents are the recipients, the grades are attributes, then this flow would be constrained by an information norm with the transmission principle of *permission* because "*[t]his information is protected under FERPA and parents do not have access to it unless the student has provided express, written authorization*" [19].

**Norm Violation**

An informational norm is breached when an action or practice disrupts the actors, attributes, or transmission principles within a given information flow. For example, a professor provides a TA with a student's confidential medical record. In other words, contextual integrity "*is* *preserved when informational norms are respected and violated when informational norms are breached*" [12].

## 2.4 CI in VACCINE

We now describe how the VACCINE-based system operationalize contextual integrity.

The actual implementation of contextual integrity involves two steps. First, the information norms need to be translated into operational privacy rules. Second, there needs to be an engine for enforcing these rules. *A system that correctly implements contextual integrity provides the following privacy guarantee: the system allows exactly those flows that adhere to the given contextual information norms.*

As a possible implementation we chose to use Datalog to formalize CI concepts as general Horn clauses and a simple *Effective Propositional Logic* (EPR) for their analysis. Our choice is largely motivated by the following reasons: Datalog is a well-studied language and formalism that has found numerous applications, including the analysis of social networks [20] and as a language for expressing privacy and security policies (see, e.g., [14, 21, 22]). We also found that Datalog and a simple EPR analysis were sufficiently expressive for our contexts of interest.

We would like to note that future implementations can opt not to use Datalog for their design of a VACCINE-based system. In fact, we consider this as one of the main advantages of the VACCINE design.

**Information flow**

The encoding of an information flow is simply a Datalog tuple:

```
(sender, recipient, subject, attribute)
```

Sender, recipient, and subject are Datalog constants used for representing actors and similarly for attribute.

**Information leakage**

The concept of norm violation in our work serves as the definition of data leakage. In other words, preventing data leakages amounts to preventing norm violations.

The CI norms describe the admissible system behaviors. As depicted in Figure 2, the CI norms are derived from a combination of external sources such as privacy handbooks and context specifications. Other external sources may also be used as a base for CI norms in a specific domain, such as legal and policy documents, professional codes of conduct, findings of empirical social scientists, privacy advocates, ethicists, stipulations of sponsoring organizations, etc.

Each CI norm specifies a set of flows. We distinguish between *actionable* and *non-actionable* norms. Actionable norms specify the operational rules that define the runtime behavior of the system. Conversely, non-actionable norms define auxiliary high-level properties that must be guaranteed by the actionable norms.

**Actionable norms.** An actionable norm explicitly specifies whether a particular type of flow is admissible in the current system state. Such a norm can be expressed directly as a Datalog rule. Our Datalog encoding uses predicates on entities that stand for contexts, actors, and attributes as described above. Central to the encoding is the predicate:

```
allowed(Ctx, Sndr, Recp, Subj, Attr).
```

This predicate models that in context `Ctx`, actor `Sndr` is allowed to send information on attribute `Attr` of actor `Subj` to actor `Recp`. For example, the following fact states that in the classroom context (denoted `class`), the flow where `bob` sends his own grade to `alice` is allowed.

```
allowed(class, bob, alice, bob, grade).
```

In order to be able to express concisely the rules, we further introduce a ternary predicate `inrole(Context, Actor, Role)`, which models that in the given context, the given actor is in the specified role. For example, the following fact states that in the classroom context, `bob` is a `student`

```
inrole(class, bob, student).
```

The next rule then codifies the actionable norm that a professor can let any student know her own grade:

```
allowed(class, Sndr, Recp, Subj, grade) :-
  inrole(class, Sndr, professor),
  inrole(class, Recp, student), Subj = Recp.
```

*Transmission principle.* In general, an actionable norm may depend on the current state of the system. As mentioned in Section 2.3, such conditional dependencies of a norm are captured by transmission principles. Here are some examples of transmission principles in the educational context:

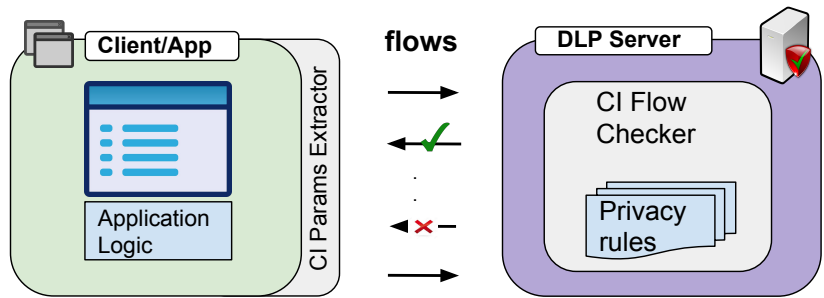

**Fig. 1.** *Runtime:* Information exchanges are processed by the CI Parameters Extractor that forms corresponding CI flows that are then fed into CI Flow Checker to check for violations with respect to existing privacy norms

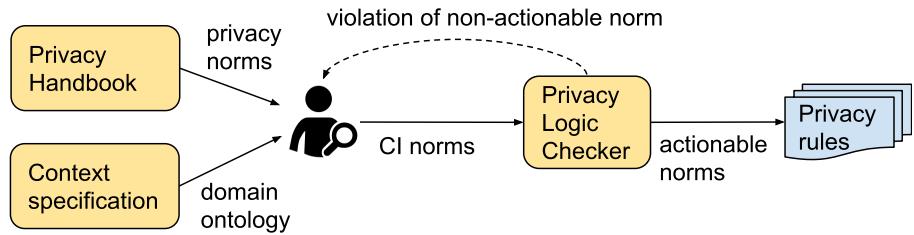

**Fig. 2.** Privacy norms are sourced offline from a combination of privacy handbooks and context specification. The norms are manually encoded into privacy logic to be verified for inconsistencies by the Privacy Logic Checker

- The receiver can access a student's academic record if he has a legitimate educational interest.
- A student's private information can be accessed or disclosed if the student gave written permission that allows such actions.
- A student's private information can be accessed or disclosed to a parent of a student if a student is in a life threatening situation.

**Temporal constraints**
We express actionable norms that involve temporal constraints in Datalog by introducing context-specific predicates that capture relevant state (e.g., whether permission to access some information has been granted in the past). Any such information on the history of the system execution that is relevant for expressing an actionable norm must be recorded explicitly in the system state. For instance, we may have a ternary predicate `lei` that encodes whether an actor has a legitimate educational interest in accessing another actor's attribute. The following Datalog fact then states that `bob` has legitimate educational interest in `alice`'s grades:

`lei(bob, grade, alice).`

The `lei` predicate can be used to encode an actionable norm that is conditioned on the transmission principle of *purpose* as stated above. Note that the definition of the predicate `lei` may change during execution of the system. E.g., the fact `lei(bob, grade, alice)` may no longer hold if `bob` leaves the university.

We also note that our EPR analysis does not reason about the actual temporal constraints (i.e., properties of sequences of information flows). We only reason about properties of individual information flows.

**Non-actionable norms.** These are properties that actionable rules need to satisfy. Some of the non-actionable norms describe allowed flows of particular importance. Other non-actionable norms describe important flows that should be blocked. More specifically, the extracted non-actionable norms fall into the following three categories:

*Implicit Norms:* These norms specify flows that are not explicitly stated in the documentation but follow from common sense and the underlying domain ontology. For instance, in the FERPA context, a student should be able to send herself a message containing her own personal data.

*Completeness Norms:* These norms also specify flows that should be subsumed by the actionable norms. If the extracted rules do not allow such flows, then there is a possible mismatch between the privacy expectations of the privacy expert and the actionable norms. For instance, one norm in the FERPA summary states that a parent can access a student's (private) non-directory information in the case that a student is in an emergency situation. At the same time, the norm does not explicitly state that a parent can also access directory information.

*Blocking Norms:* These norms specify flows that should be blocked by all means.

# 3 Architecture

As depicted by Figure 1, systems relying on VACCINE follow a client-server architecture:

**Client** The client side is designed to analyze communication exchanges in order to extract the relevant information flows, i.e., CI parameters. We envision two main types of operational deployment setups: a) as a standalone application b) as an addon to an existing application such as email or a browser.

**Server** The server is located in the local (enterprise) network and serves as a trusted entity to maintain the privacy logic. Depending on the preferred setup, the server can securely push the logic to the client side which will perform the check on the extracted information flows or let the client side send the extracted flows for checking to the server.

The operational flow of VACCINE used to build powerful DLP systems with the above described architecture comprises two main phases:

*1) Offline* As illustrated in Figure 2, a user of VACCINE first extracts the privacy logic in the form of CI norms from relevant external sources. Then, VACCINE uses formal verification techniques to inform the user in a feedback loop of any inconsistencies within the extracted norms. At the end of this process, the consistent norms are automatically converted into actionable privacy rules.

*2) Runtime* As illustrated in Figure 1, during runtime VACCINE mediates the flow of information between participants to enforce the privacy rules that have been extracted in the offline phase. Information exchanges are processed by the CI Parameters Extractor provided by the application to generate corresponding information flows that are checked against the privacy rules by the CI Flow Checker.

## 3.1 Extracting Privacy Logic

The process of extracting privacy logic consists of extracting actionable and non-actionable norms from a privacy handbook and checking their consistency. By consistency, we mean that actionable and non-actionable norms agree. For instance, no flows defined by blocking norms should be allowed by actionable norms.[1]

In practice, this involves the identification of the required auxiliary state predicates and the actual specification of the actionable norms in Datalog as described in the previous Section. The user may also identify non-actionable norms that should be guaranteed by the system but that may not be expressible in Datalog. VACCINE supports the automated static verification of such non-actionable norms as long as they remain expressible within a decidable logic. This functionality is realized by the Privacy Logic Checker.

We also point here to few other properties of such privacy logic model. First, our model enjoys the whitelisting feature: only flows defined by actionable norms are allowed and nothing else, instead of explicitly denoting what flows are not allowed. Further, since checking flow admissibility is done using a Datalog interpreter, as explained later in this section, the correctness of well-studied Datalog semantics and the consistency of the privacy rules make the resulting DLP system correct-by-construction, subject to the extracted norms. We now explain the technical details behind checking the consistency of the extracted norms.

### 3.1.1 Privacy Logic Checker

The Privacy Logic Checker serves a tool to assist a DLP team in verifying CI norms, improving their reliability and overall understanding of a particular privacy context. In VACCINE, this process amounts to checking whether actionable and non-actionable norms are in line, and is performed *statically* before the rules for the actionable norms are in effect. In case this check fails, i.e., there is an inconsistency between actionable norms and non-actionable norms, the norms need to be corrected. This process is repeated until there are no more inconsistencies as depicted by the feedback loop in Figure 2.

---

[1] We note that this notion of consistency should be not confused with the notion of consistency in formal logic.

**Under the Hood of Privacy Logic Verification**
The problem of verifying that a given set of actionable norms $\mathcal{R}$ ensures a given non-actionable norm $P$ amounts to checking logical validity of the implication $\mathcal{R} \Rightarrow P$, or dually, that the conjunction $\mathcal{R} \wedge \neg P$ is unsatisfiable. If the negation of $P$ is expressible in Datalog, the latter can be checked directly using Datalog queries. However, in general, Datalog queries are not sufficiently expressive to verify complex high-level properties. Fortunately, we can embed Datalog in a more expressive logic that is still amenable to automated reasoning and yields tractable performance for the static verification of high-level properties in practice.

Datalog is a fragment of first-order predicate logic. Specifically, suppose we are given a set of rules $\mathcal{R} = \{R_1, \ldots, R_n\}$ where each rule $R_i$ is a Datalog clause of the form

$$\texttt{allowed}(C, Sn, R, Su, A) \text{ :- } C_{i,1}, \ldots, C_{i,m_i}.$$

and the atoms $C_{i,j}$ are either `in_role` and context-specific predicates over the variables in the head of the clause or equalities between these variables and constants such as `student`, etc. Then the semantics of these clauses is captured by the following quantified formula:

$$\forall C, Sn, R, Su, A.$$
$$\texttt{allowed}(C, Sn, R, Su, A) \Leftrightarrow$$
$$(C_{1,1} \wedge \cdots \wedge C_{1,m_1}) \vee \cdots \vee (C_{n,1} \wedge \cdots \wedge C_{n,m_n})$$

This formula falls into EPR, a decidable fragment of first-order predicate logic [23]. Several automated theorem provers implement decision procedures for EPR, e.g., the Satisfiability Modulo Theories solver Z3 [24]. If the negation of the non-actionable norm $P$ is also expressible in EPR, then we can use Z3 to automatically check that the rules $\mathcal{R}$ guarantee $P$. Fortunately, many properties of interest are indeed expressible in EPR. For example, the following EPR formula expresses that in the classroom context, a professor should not be allowed to send a student's grade to any other student, unless that other student is a TA:

$$\forall Sn, R, Su.\ \texttt{in\_role}(\text{class}, Sn, \text{professor}) \wedge$$
$$\texttt{in\_role}(\text{class}, R, \text{student}) \wedge$$
$$\texttt{allowed}(\text{class}, Sn, R, Su, \text{grade}) \Rightarrow$$
$$Su = R \vee \texttt{in\_role}(\text{class}, R, \text{TA})$$

Note that this property cannot be checked with a simple Datalog query. We need the additional expressiveness provided by EPR. More generally, EPR can express properties about transitive information flows that involve arbitrarily long sequences of information exchanges.

Our encoding into EPR of norms and properties remains within a decidable logic that admits practical decision procedures. This means that we can verify these properties fully automatically using tools such as Z3. In particular, a failed verification attempt is always due to an actual violation of a non-actionable norm (as opposed to an incompleteness in the verification approach). If a particular non-actionable norm is violated, the theorem prover will produce a model describing a sequence of information flows that respects the rules but violates the norm. Using this model, we can then identify the rules that are responsible for the violation.

### 3.2 CI Flow Checker

Checking whether a flow complies with the privacy logic amounts to performing a single query of the `allowed` predicate.

Going back to our educational privacy context of classroom, denoted by `class`, suppose that our state of the system includes the following facts:

```
inrole(class, bob, student).
inrole(class, alice, student).
inrole(class, steve, professor).
```

Then we can use the query mechanism provided by a Datalog interpreter to check whether a specific information flow satisfies all the extracted privacy rules. For instance, given the above facts and the privacy rule described in Sec. 2.4, the query

```
?- allowed(class, steve, bob, bob, grade).
```

evaluates to `true`, indicating that the corresponding information flow is admissible. The query

```
?- allowed(class, steve, alice, bob, grade).
```

evaluates to `false`, indicating that this flow is not permitted. The semantics of Datalog guarantees that no norm-violating flows will be permitted at runtime.

### 3.3 CI Flow Extraction

To identify the relevant CI parameters we reference the domain ontology for existing actors and a range of possible attributes. This data needs to be provided to the CI Parameters Extractor to identify the actors and attributes within an information exchange and construct corresponding CI flows. The exact way in which flows are extracted from an information exchange depends on the actual exchange platform. However, we now sketch common information exchange patterns and describe their conversion to CI flows.

*Explicit:* This represents an information exchange between a sender and an explicit recipient, which for instance, can be formed by an email, instant message, SMS, etc. Each time a user explicitly states a recipient of a message, e.g., in an email in TO/CC/BCC fields, it constitutes a CI flows from the sender to explicitly specified recipient. Explicit information exchange lends itself well to CI flows, constrained by respective transition principles.

*Implicit:* Implicit information exchanges represent interactions where the recipient was not explicitly specified. These interactions are formed as a byproduct of the application logic. For example, when a user mentions someone in a tweet, all the followers of that posting user will also see it as well, even without explicitly being addressed as a recipient. Search functionality also falls under this category because the posted information reaches beyond the explicitly intended recipients. Authentication and authorization mechanisms will typically comprise multiple interactions that stem from one explicit information exchange, e.g., to login into a system. An implicit information exchange yield an additional CI flow with same attributes and subject between the new sender and the recipient. For example: suppose Alice sent her grade to Bob in the first email. Bob wants to forward that as an attachment in his email to Mary. To check whether there is a violation, we need to check two resulting CI flows, one with parameters of Bob email to Mary and the (implicit) second with CI parameters from the attached email, i.e., checking whether "Bob can send Alice's grade to Mary".

*Transitive:* A transitive information exchange emerges when a recipient forwards the information to a different recipient, and so on. For example, when an email is being forwarded, a tweet retweeted, etc. This is a great example that shows how the CI theory differs from previous approaches. In theories that rely on private/public dichotomy, a transitive information exchange can be problematic in cases where private information is exchanged between A and B, then forwarded by B to C, and there is a privacy rule saying that A cannot share that information with C. However, with CI this is less problematic because it looks at each flow separately and only in the context of the CI parameters. For example, let us assume a doctor shares with Jon his test results. This is highly confidential information, which the doctor cannot share without Jon's explicit consent. So if the doctor shares the information with Jon's employee, this will violate the privacy norm. However, nothing precludes Jon from sharing his personal medical records with another actor. From the CI perspective, the resulting transitive exchange from A to C is irrelevant as such. Only the individual flows from A to B and B to C are relevant for the privacy logic.

Importantly, each action defined by an application logic typically comprises several of the above described information exchange patterns, each possibly yielding multiple CI flows.

*Information exchange complexity.* The complexity of information exchanges will often depend on a particular application logic. Nevertheless, we give a general worst case scenario analysis independent of any application. Suppose we have a sequence of $n$ related information exchanges, such as an email thread or subsequent retweets. The complexity of an information exchange sequence ($CIES$) is the sum of complexities of each constituting information exchange ($CIE$). The complexity of a single information exchange, $CIE_i$ in the sequence is defined in terms of the CI parameters in the generated CI flows.

$$CIES = \sum_{i=1}^{n} CIE_i = \sum_{i=1}^{n} R_i \times A_i \times S_i \qquad (1)$$

Here we denote by $A_i$, $R_i$, $S_i$ the number of attributes, recipients, and subjects associated with flows generated from the information exchange $i$. We measure the performance of our systems under simulated information exchange complexities in Section 4.3.

**Proof-of-concept implementation**
We have implemented a proof-of-concept DLP system for educational setting using VACCINE. The client side of the system is a Chrome browser add-on named Contextual Integrity-based Gmail App (CIGApp), whose screenshot is shown in Figure 3. The CIGApp extension assists users by retaining the integrity of the information exchanges generated as a result of their emails by ensuring that the subsequent email exchanges are consistent with the FERPA context [19]. For our prototype, we used the InboxSDK [25] library, which is widely used by major companies such as Dropbox and Stripe, to take control over the user's Gmail inbox and extracts relevant CI flows. We chose Gmail as our target application layer since it is widely adopted by the majority of top 100 universities [26].

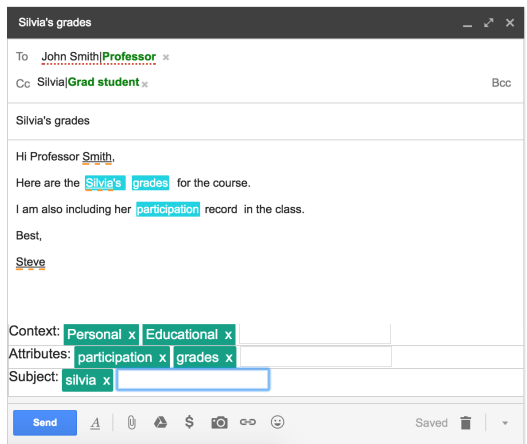

**Fig. 3.** Screenshot from the proof-of-concept prototype with relevant CI parameters identified

The DLP server is hosted on the Heroku cloud service [27]. It hosts privacy rules encoded using Datalog and SWI-Prolog execution engine that can interpret Datalog (and Prolog) programs [28]. The interpreter is used for checking compliance of flows against the rules. The actual rules and their creation are explained in the evaluation part of this section.

## 4 Evaluation

Our evaluation examines the applicability, feasibility, and scalability of VACCINE DLP system in two privacy contexts: the educational context as captured by the FERPA regulation [19], and an enterprise context in the Enron corporation [29–31]. We primary chose the educational context because it is the one with which we are most familiar but also to illustrate that our framework will allow building regulation driven DLP-type systems which are currently missing in such (specialized) contexts. We note that VACCINE can also be used with other contexts e.g., in enforcing the Health Insurance Portability and Accountability Act (HIPAA) Privacy Rule [32]. Finally, in the absence of a real world email dataset from an educational institute, we use an Enron email corpus to simulate the volume exchanges in a large enterprise to test the feasibility of our approach.

Our experiments focus on evaluating the core components of the VACCINE framework. Specifically, we aim to answer the following questions concerning the DLP server side:

1. How formal methods can assist in the creation of a consistent set of privacy rules?
2. What is the expected server load in terms of generated flows from information exchanges?
3. How efficiently can CI flows be checked against the privacy rules?
4. What is the network performance of our prototype?
5. How effective is the VACCINE framework in preventing potentially unauthorized flows in a real-world emulated context?

### 4.1 Methodology

Our CI encoding of FERPA is based on a summary of the FERPA law applied to a university context [33]. The document summarizes how personal information related to students (directory information, grades, etc.) can be accessed and shared. In the document, we have identified 12 roles, 5 attributes, and 3 transmission principles that define the space of potential CI flows in this context. We chose the Enron corporation for our enterprise context since both its code of ethics [29] as well as a 1.7GB big data set consisting of hundreds of thousands of emails sent within the company are publicly available [30, 31]. From Enron's code of ethics we extracted CI parameters for an enterprise context. Our focus was on CI flows related to the exchange of sensitive personally identifiable information (PII) such as passwords, bank account information, phone numbers, passwords, etc. We identified 12 roles, 6 attributes, and 3 transmission principles that define the space of CI flows in this context. In experiments that involve the creation of CI flows from the Enron email data set, we have mapped the names and email addresses of employees to their roles in the Enron corporation by using the results of a prior analysis of this data set [34].

Our evaluation primarily focuses on the server side of our implementation as that is the product of using VACCINE to create a prototype DLP system for educational settings. Unless otherwise noted, all of the experiments were carried on a `Intel(R) Xeon(R)` machine with eight 3.6GHz cores.

### 4.2 Experiences with Privacy Rules Consistency Verification

In our first experiment, we checked if formal representation VACCINE uses for privacy norms is effective at aiding the creation of a consistent privacy logic for a given context. To this end, we extracted CI norms from the privacy documentations for the FERPA and En-ron context (in practice this is done by a DLP team) and checked their formal consistency using the theorem prover Z3 (this process can also be automated), as described in Sec. 3.1.1. In both cases, we extracted a set of *actionable norms* that explicitly state which flows are allowed in the context and *non-actionable norms* that express consistency properties of the actionable norms, as discussed in detail in Section 2.4. The verification ensures that the actionable norms enforce the non-actionable norms. If the actionable norms disagree with some non-actionable norms, we say that the norms are inconsistent.

**Results for FERPA** We manually extracted privacy norms from the FERPA summary by following the process described in Figure 2. Eventually, we extracted 28 FERPA norms of which 18 are actionable (rules) and 10 non-actionable (Section 2.4). It took three iterations of the check-refine loop in Figure 2 to obtain a consistent set of actionable norms. The results of the experiment are summarized in Table 2. We show the number of norms for each non-actionable norm class, the average time taken to check the consistency of these norms, and the number of consistency checks that passed in each iteration of the process. As can be seen, Z3 performed the checking instantaneously.

In the first iteration of the norm creation process, we extracted an initial set of 15 actionable norms and checked whether they are consistent with the 3 implicit norms, 3 completeness norms, and 4 blocking norms also extracted from the FERPA summary. Checking each of these non-actionable norms was done using a single query to the theorem prover. As it turned out, none of our implicit norms were consistent with the actionable norms. Therefore, we added three additional actionable norms that allowed the flows defined by our implicit norms, resulting in a total of 18 rules.

In the second iteration, we rechecked the consistency of the 18 rules subject to the completeness and blocking norms. We first observed that one completeness norm was still violated. This norm states that a parent should be able to access student's (private) directory information in the case of an emergency. Our rules blocked such flows although we expected them to be allowed: if parents are allowed to see student's highly-sensitive private information in the case of emergency, then they should be able to also access less sensitive directory information. We then reached out to the authors of the FERPA summary from which we extracted the norms. According to the authors, they believe that parents' access to student's directory information in emer-gency situations is implied, which was not clear from the text in the handbook. We slightly modified one actionable norm to also account for this omission, i.e., to allow parents to access student's directory information in the case of emergency. Also, our rules did not imply one of the blocking norms, which states that parents should not be able to access student's private information. The rules do block such flows except in certain cases where parents have written access permission to do so. We believe the authors of the summary document also implicitly excluded these exceptional cases when formulating the blocking norm. We therefore modified the corresponding non-actionable norm accordingly. In the final iteration of our check-refine loop, all the rules were consistent with the non-actionable norms.

|  | checks per iter. | consistent I | II | III | avg. time |
|---|---|---|---|---|---|
| **Implicit** | 3 | 0 | 3 | 3 | 0.01s |
| Completeness | 3 | 2 | 2 | 3 | 0.01s |
| Blocking | 4 | 3 | 3 | 4 | 0.01s |

**Table 2.** Results of checking inconsistencies between actionable and non-actionable rules in iterations I-III

**Results for Enron** For this context, we created 43 Enron privacy rules that focus on access and disclosure of PII in a corporate setting. The norm creation process was informed by the available documentation of Enron's code of ethics and its organizational structure. However, due to the lack of a publicly available (Enron) enterprise privacy handbook that specifies precise norms, as in the case of FERPA, we created norms to enforce a simulated *ethical wall* on email communication between Enron employees. An ethical wall [35] refers to establishing an environment that will prevent information exchange between specific departments within the organization to avoid any conflicts of interest.

Our rule creation process focused on each type of PII and how agents in the various enterprise roles can exchange the corresponding attributes. A typical example of a rule is: *Enron employees can send their social security number (only) to administrative managers*. We did not have to create any implicit, completeness, or blocking norms for this context. We immediately converted implicit and completeness norms into actionable rules. This is a by-product of our initial experience that we gained from analyzing the FERPA norms and the fact that we were given more freedom in creating actionable norms for the Enron context. The case of block-ing norms is more interesting. We created the rules by following our white-list methodology that every flow is disallowed by default unless there exists a rule that explicitly allows it. Initially, we had no rules, thus disallowing any flow whatsoever. We subsequently added the rules allowing more and more flows. Consequently, we avoided the need for creating blocking norms in the first place since they are implicitly guaranteed by our approach. This experience enforced our belief that the VACCINE privacy model could be the guiding principle for writing privacy handbooks.

**Summary** We conclude that contextual integrity and verification techniques allow one to detect flaws in the privacy model and subsequently eliminate these flaws by refining or adding new privacy norms.

## 4.3 DLP Flow Load

The goal of the next experiment was to assess the server load that a DLP system created using VACCINE framework could experience in practice. Specifically, we measured how many CI flows are typically generated in an email conversation. For this purpose, we used the Enron email data set [30, 31].

**Setup** We grouped the emails in the Enron data set into conversations. We define an email conversation as a set of emails with the same email subject (modulo reply and forwarding prefixes, etc.). We extracted from each email a set of quadruples (one per recipient) consisting of the sender email address, a recipient email address, the time the email was sent, and the recipient type which can be TO, CC, and BCC, omitting the email content and other metadata. Here is an example of such a tuple:

(tracy.geaccone@enron.com,
 sarah.taylor@enron.com, 2001-11-27 12:48:05,TO)

We identified every unique email address with a unique participant. For emails sent to mailing lists, we simply simulated broadcasting the email to all of the active conversation participants.

Out of the whole Enron email data set, we randomly chose 2,000 email conversations involving approximately 80,000 emails and 11,500 participants in total, as shown in Table 3 (first and second row). We used the email time and recipient type fields to recreate email threads for each conversation. There are two email threads per conversation on average, as shown in Table 3 (third row). Due to some missing information in the data set, we observed cases where participants send emails in the middle of the conversation although they never received any emails in the conversation up to that point. In our analysis, such emails start a new conversation thread, which explains multiple threads per conversation.

|  | Total | Min | Avg | Max |
|---|---|---|---|---|
| **Emails** | **78,386** | **1** | **39** | **15,555** |
| Participants | 11,609 | 1 | 17 | 2,428 |
| Threads | 4,082 | 1 | 2 | 196 |
| Flows | 218,070 | 1 | 109 | 85,280 |

**Table 3.** Statistics for each of the 2000 Enron email conversations

We used the extracted email threads to generate CI information flows for each conversation. As mentioned in Section 3, a single email can generate multiple flows between different participants. This is because an email often quotes other emails that have been sent earlier in the same conversation. In such cases, the new email also generates transitive flows involving subjects and attributes from the quoted emails to the new recipients. To model the worst case scenario in terms of the information flow complexity, whenever an email is sent by some sender in a conversation, we assume the sender creates new content and also transitively sends all of the emails in the conversation up to that point. Also, a CI flow has a subject and attribute. For the purposes of this experiment, the actual subject and attribute values are not important. For transitive flows, however, the subject is the sender of the corresponding email that is being quoted.

**Results** Given the above dataset, we measure how many flows an email yields in practice to asses the load stress on the server. As it can be already inferred from Table 3, a single email yields 2.78 flows on average. Moreover, Figure 4 shows the frequency distribution of emails depending on the number of flows they generate. It can be seen that the flow-per-email ratio is in general low, reaching 15 at maximum. This suggests that the load on a DLP server created using VACCINE will be tolerable in practice. This is further evidenced by Figure 5 showing a linear increase in the number of generated flows subject to total number of thread emails.

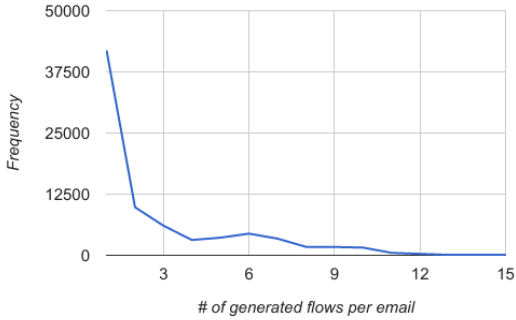

**Fig. 4.** Email frequency distribution based on the number of generated flows per email

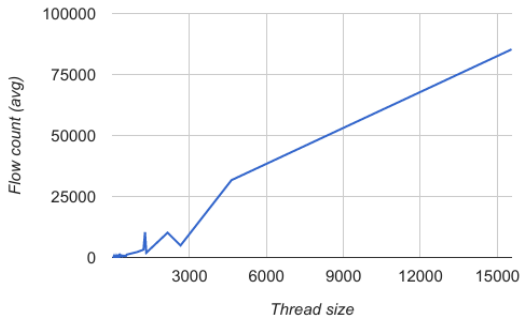

**Fig. 5.** Average number of generated flows per size of the conversation thread

### 4.4 Flow Checking Performance

The purpose of our next experiment was to measure the general performance of checking flows against privacy rules. For that purpose, we reused the extracted consistent Enron actionable norms that we created in the experiment described in Sec. 4.2. We manually converted these norms into Datalog (privacy) rules; this process can also be automated. The rules together with the Datalog engine constitute the Enron CI flow checker (see Section 3). We chose the Enron context instead of FERPA for this experiment as it aligns with the email data set from which we extract CI flows.

**Setup** We created a state of the system by assigning roles to Enron employees. For 138 Enron employees we used their actual company roles taken from publicly available data sources [36]. For the remaining employees whose actual roles are unknown, we assigned roles randomly. We also initialized the system state using reasonable assumptions about disclosure privileges for PII information, namely that within each department (i.e., role) employees can disclose each other's phone numbers and addresses. The transmission principles of certain norms depend on these disclosure privileges. The resulting system state comprises about 7500 Datalog facts, 2MB in total. This state of the system can be thought of as a database that is queried by the server in order to check flows against the privacy rules. We keep the database in server memory for better performance.

**Results** We used the flows generated from the Enron email data set to test the speed of checking flow admissibility. More precisely, for each Enron flow we randomly picked a sender or recipient as the subject of the flow. We also randomly picked attributes. *Datalog responsiveness* Our first experiment set to test the responsiveness of the Datalog interpreter itself. We automatically submitted the generated flows one after the other to the interpreter and measured the response times. The results of the experiment are shown in Table 4 (first row). As it can be seen from the average time of processing a single query, Datalog flow admissibility checking is very efficient. This is reflected in a good throughput.

|  | Total time | Time per flow | Allowed flows | Rejected flows |
|---|---|---|---|---|
| **non-mutable** | **206s** | **0.9ms** | **51,937** | **166,133** |
| state updates | 358s | 1.6ms | 51,911 | 166,159 |
| rule updates | 616s | 2.8ms | 51,898 | 166,172 |

**Table 4.** Results of checking 218,070 flows against the Enron rules

In a typical system of an enterprise network, a common information exchange will often also mutate the state of the system. For instance, employees might change their roles as they are being promoted in the company. Hence, we also tested the responsiveness of the flow checker in a setting where the state is updated in between incoming flow admissibility queries. We ran the exact same experiment, except that each query was followed by a random state update request. The experiment results are shown in Table 4 (second row).

Finally, we wanted to see how flow checking times change if we dynamically add rules to the privacy logic engine. In practice, rules might be added to the system as the privacy norms evolve over time. Hence, we repeated the previous experiment where in every 10th update, we updated the database with a random rule rather than with a random state change; hence, the system had more than 20,000 rules at the end of the exper-iment. Each randomly generated rule either (1) allows flows with a random attribute where the sender, recipient, and the subject need to satisfy some Enron transition principle or (2) allows a concrete random specific flow. The results are shown in Table 4 (third row). Although the average time to check a flow increases with the increasing number of privacy rules, the server responsiveness is still very high. We note that the results include the round trip time taken by a query to check whether a given flow is allowed. *Network deployment* To test the general VACCINE server responsiveness and network performance, we used the free Heroku dyno type service [37] which provides a server with 512MB RAM, 1024MB swap, and 8 (virtual) CPU cores (Intel(R) Xeon(R) CPU E5-2670 v2 @ 2.50GHz). Using this setup, we reran the same experiment as above (with no rule and state updates) but this time sent all the flows in parallel instead of sequentially. This way we tested the performance of the Datalog privacy engine more thoroughly in a realistic setting. Figure 6 shows that the connection time for each request averaged 3 milliseconds and the service time $2.07$ milliseconds. The average size of a request was $158$ bytes.

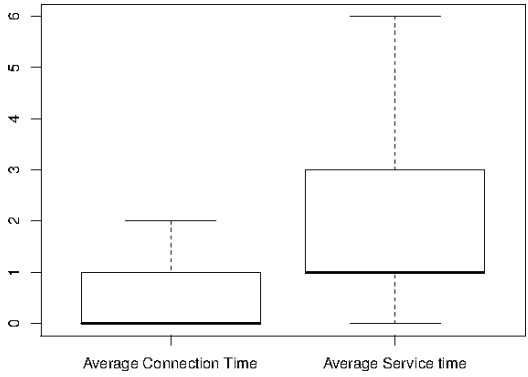

**Fig. 6.** Connection and service times. Mean connection time is 1.34ms, service time 3.2ms, and mean request size is 158 bytes.

Our results indicate that a typical server will be able to process requests efficiently without incurring a significant overhead onto the network.

### 4.5 Data Leakage Prevention

In the final experiment, we analyzed how effective VACCINE is at preventing the leakage of privacy sensitive information in a realistic setting.

**Setup** We extracted CI flows from the Enron email data set [30, 31]. As there are 138 Enron employees with known company roles, we only considered email exchanges between those employees as well as a single generic third party agent that we associated with all external email address. From each email, we extract PII attributes using a simple keyword pattern matching mechanism. Our flow extractor for this experiment simply assumes that the sender is the subject of a flow, except when the sender is the generic third party agent, in which case the subject is the recipient. This exception simulates the adversarial scenario where a third party asks for, or generally refers to, sensitive personal information of some Enron employee. We chose this simple approach as more complete solutions require the use of more advanced techniques, which is out of the scope of this paper. In total, we extracted 592 flows for this experiment.

**Results** 39 flows were rejected by the engine for the extracted Enron rules. Interestingly, only password (37) and bank account information (2) appear as attributes in the rejected flows. In most instances, the rejected flows with *password* as the attribute stem from emails where an employee forwards another email of an external website that asks for a password update, and when employees send their password information to a third party personal email address. We also observed the case when one employee sent her own password for accessing Enron's internal network to another coworker. The two rejected flows involving the bank account attribute are actually benign. In all such cases where a flow was falsely rejected, the misclassification was caused by our flow extractor rather than the flow checking engine.

| role | sender | recipient |
|---|---|---|
| CEO | 0 | 3 |
| president | 4 | 4 |
| vice president | 13 | 0 |
| director | 2 | 2 |
| lawyer | 0 | 1 |
| trader | 1 | 0 |
| manager | 9 | 0 |
| employee | 10 | 2 |
| third party | 0 | 27 |

**Table 5.** Breakdown of the number of rejected flows based on sender and recipient role

For a more detailed analysis of the rejected flows, we categorized them based on the role of the sender and recipient. For each role, we show how much flows were rejected when the sender (resp. recipient), was in that role. The results are shown in Table 5. For instance, there was not a single rejected flow where the CEO was a sender, but 3 flows were rejected when a CEO was the recipient. As can be seen, the majority of the rejected flows have a third party agent as the recipient, which is consistent with our manual inspection of the corresponding emails.

The flows that the engine classified as admissible mostly have bank account information (291) and passwords (164) as attributes. We manually inspected 200 of these flows to see if there are any potentially malicious flows that were allowed by our rules. We found one such case where a vice president was sharing contact information about another higher rank employee to employees of lower rank. Our rules prohibit such information sharing; we allow contact information sharing only with recipients who are further up in the company hierarchy than the sender. The reason why this flow was allowed is because our mechanism extracted the wrong subject for this flow. The extracted flow itself was correctly classified. This problem could be rectified by using an advanced subject extraction mechanism. All the other accepted flows that we inspected were extracted correctly.

These experiments show that VACCINE provides an effective tool for data leakage prevention, modulo an effective flow extractor.

# 5 Related Work

The key building block of the VACCINE framework is to define privacy and confidentiality specifications of users and enterprises in the language of Contextual Integrity (CI). We already discussed how VACCINE compares to existing commercial DLP solutions in Section 2. The research community has explored sophisticated email DLP methods mostly relying on a combination of machine learning and related classification techniques. These approaches roughly fall under content- and behavior-driven methodologies [1].

**Content-based approaches** [9, 38–40] determine the sensitivity of a message based on textual content analysis, either through keyword, regex pattern matching, or machine learning techniques.

**Behavior-based approaches** [41–43] analyze the organizational structure, common roles and duties, as well as past interaction to identify "normal" behavior patterns and use them to detect unlikely information flows between a pair of sender and recipient.

Our work is largely orthogonal to these approaches. While prior solutions explored advancing techniques in identifying insensitive data and malicious behavior patterns, VACCINE framework offers a robust way to express, verify and administer DLP policies.

Here we focus on related work in CI-based formalization for building privacy-aware tools.

Defining privacy [12] in terms of CI is now a well-established approach in the privacy and computer science research communities. It is used to describe contextual informational norms, to detect infractions of these norms, and in approaches to accountability and enforcement [13, 44]. Much of the prior work in this space has focused on sophisticated formal logics for describing CI norms that involve complex temporal properties of sequences of information flows such as those involved in HIPAA [45–48]. However, these logics are typically too expressive to serve as a suitable foundation for tools that mechanize reasoning about privacy norms in real-world systems. Our observation is that in the DLP setting, the much simpler logical framework provided by Datalog and EPR is sufficiently expressive to express the relevant norms.

Other relevant works include [49] where the authors proposed a prototype messaging system that checks whether information exchanges comply with HIPAA in a medical domain. The proposed approach uses pLogic, a stratified version of Datalog, to formalize HIPAA regulations related to information sharing. This work is similar to our approach, it is based on a different CI model that introduces explicit additional parameters such as *purpose* and *believe* which are not part of the theory of CI and specific to HIPAA. Also, our work goes beyond checking information flows against policies, as VACCINE also ensures that the policies themselves are logically consistent. In [44], the authors propose computational and information models of Implicit Contextual Integrity in a social network. Our work complements such efforts with a framework that provides stronger guarantees in checking that information exchanges in the system follow the norms and verifying them for any inconsistencies with respect to privacy policies within a particular domain.

# 6 Conclusions

We presented VACCINE, a framework for building DLP systems. The key ingredients of VACCINE are the use of the language of Contextual Integrity to define (enterprise) DLP privacy policies and their formalization in verifiable logic. In our evaluation, we have demon-strated how an administrator can leverage the VAC-CINE framework to derive a CI-based logic and then use it to prevent unintended information leakage within the organizational network, and beyond.